\documentclass[twocolumn,aps,pra,showpacs,groupedaddress]{revtex4-1}
\usepackage[matrix,frame,arrow]{xy}
\usepackage{amsmath}

\usepackage{epsfig,amssymb,amsmath,mathrsfs,color} 
\usepackage[active]{srcltx} 
\usepackage[hypertex,linkcolor=red]{hyperref}
\usepackage[matrix,frame,arrow]{xypic}
\newtheorem{lemma}{Lemma} \newtheorem{proposition}{Proposition}
\newtheorem{corollary}{Corollary} \newtheorem{theorem}{Theorem}
 \newtheorem{definition}{Definition}

 \def\>{\rangle}
\def\<{\langle}         \def\rX{{\rm X}} \def\rY{{\rm Y}}

\begin{document}
\title{Information theoretic axioms for Quantum Theory}
\author{Marco Zaopo}\email{marco.zaopo@unipv.it} 
\affiliation{Dipartimento di Fisica, Universit\`a
  di Pavia, via Bassi 6, 27100 Pavia, Italy}
\email{marco.zaopo@unipv.it}
\date{\today}
\begin{abstract}

In this paper we derive the complex Hilbert space formalism of quantum theory
from four simple information theoretic axioms. It is shown that quantum theory
is the only non classical probabilistic theory satisfying the following axioms:
distinguishability, conservation, reversibility,
composition. The new results of this reconstruction compared to
other reconstructions by other authors are: (i) we get rid of
axiom ``subspace'' in favor of axiom conservation eliminating
mathematical requirements contained in previous axiomatics; (ii) we are able to classify all the probabilistic theories that are
consistent requiring (a) only the first two axioms (b) only the
first three axioms; this could be useful in
experimental tests of quantum theory since it gives the possibility
to understand whether or not other mathematical models could be
consistent with such tests; (iii) we provide a connection between two different
approaches to quantum foundations, quantum logic and the one based on
information theoretic primitives showing that any theory satisfying the
first two axioms given above either is classical or is a theory in
which physical systems are described by a projective geometry. 
\end{abstract}
\maketitle

One of the most curious facts about quantum theory is that it explains almost every physical phenomenon except
from gravity and it is still not clear why it works so well.
The first attempts to give rigorous foundations for the rules of
quantum theory initiated a subject called quantum logic
\cite{ql1,ql2,ql3,ql4,ql5,ql6,ql7,ql8}. The starting point in quantum
logic is that propositions related to measurements performable on a quantum
system can be associated to sentences of a propositional
calculus. When the system is classical the propositions related to
classical measurements form a Boolean algebra and Boolean
algebras are the algebraic models of the calculus of classical
logic. The main question that quantum logic adresses is: when a
Boolean algebra is relaxed into an orthomodular nondistributive
lattice (i.e. a generalization of the lattice of subspaces of
a hilbert space), which logic is it the model of? ``Quantum Logic'' is
the name that designates the answer but there are several views about
the content of this name and its physical significance \cite{beltra}.  
Other notable and conceptually simpler ways to look for a physical
explanation of the rules of quantum theory is to consider
hidden variables models \cite{Bohm, toft,
  ghira}. The starting
point to formulate these models is that the state of
a quantum system describes an ontic property of the system and the
randomness in quantum experiments is simply a consequence of
mediating the result on many repetitions of the experiment. These models are
appealing because they give simple explanations to questions regarding the onthology of the state
of a quantum system or the nature of the measurement process. In such models a
physical system infact always possesses a definite value of a physical
quantity and
a measurement is just a read of this value. One of the main drawback
to consider this as the physical explanation of quantum phenomena is
that hidden variables in these models cannot be Lorentz invariant and, at the same
time, be used in a model equivalent to quantum theory from the predictive point of
view \cite{hardy1}. The advent of quantum information theory opened a new direction in the
research of foundations of quantum theory. Quantum information
showed that the controversial physical phenomena arising from the mathematics
of quantum theory, can be exploited for
information theoretic tasks that are impossible in a world governed by classical
physics \cite{tele,bb84,wiesner,e91}. It is then natural to ask wether it is
possible to put as foundations of the mathematical rules of quantum theory
a set of information theoretic principles. The first work
that partially answered to this question in a positive way was \cite{axioh}. Almost ten years
after the appearence of that paper, finding informational principles
for quantum theory is becoming a quite active area of research \cite{goyal,axiom,axiob,axioq,fivel,axioh1}. Based on
the reconstruction of \cite{axioh}, in \cite{axiob} it is given an
argument to eliminate the use of one of the axioms in the first
reconstruction, Simplicity.
In \cite{axiom}, the argument developed in \cite{axiob} is used to derive a new set
of requirements that, if imposed to a probabilistic theory, are equivalent to the mathematical formalism of
quantum theory. In \cite{puri} it is considered
purification as a foundational principle at the base of all the new
information theoretic features of quantum information. It is
shown that all theories satisfying purification (namely all states of
a system A
are in one to one correspondence with pure states defined on a larger
composite system
AB) and local discriminability
(that is shown to be equivalent to local tomography) are
very similar to quantum theory from an informational point of view. Starting from this
work the same authors gave in \cite{axioq} a reconstruction of quantum theory
from six informational axioms, two of which are
purification and local discriminability. The author of \cite{axioh}, ten years after his first
reconstruction, invented a formulation of quantum theory based on
a new mathematical object called Duotensor and gave a new set of
operational/informational principles that are proved to be equivalent
to this formulation of quantum
theory \cite{axioh1}.

In this paper we give a new reconstruction of quantum theory based on
a new set of informational principles. The main result is the
following: the only non classical probabilistic theory satisfying a list of four
axioms - Distinguishability, Conservation,
Reversibility, Compositon - is quantum theory. The new results of this reconstruction compared to
other reconstructions by other authors are: (i) we get rid of the
subspace axiom (see \cite{axioh, axiom, axiob}) and of any other different formulation of it (see the
compression axiom in \cite{axioq}) (iia) we show that every
probabilistic theory in which axioms distinguishability and
conservation hold is either classical or a probabilistic theory in
which pure states are points of a projective space defined over a
generic field of numbers; this is a generalization of quantum theory
in which the superposition principle holds with amplitudes not
necessarily complex but belonging to a generic field of numbers (iib)
we show that if in (iia) is required axiom reversibility then the
elementary system must be a hypersphere in generic $d$
dimension; among these theories we find generalizations of quantum
theory in which complex numbers can be substituted by
normed real division algebras (i.e. Reals, Complex, Quaternions and
Octonions) (iic) we argue that quantum theory over reals is consistent
with all axioms a part from local tomography of axiom
composition (see below) that must be subsituted with 2-local tomograhy
(see \cite{Wootershardy}). These classifications could be useful to check whether the predictions of
a given experimental test of quantum theory are in principle consistent with other
probabilistic theories or not.  

The paper is organized as follows. In section \ref{pt} it is presented the framework of probabilistic
theories from which it is reconstructed quantum theory and in section \ref{aqt} they are presented the axioms.
With section \ref{mqt} it begins the derivation. Our derivation is based on the fact that pure states of a quantum
system with dimension $n$ are
points of a complex projective space of dimension $n-1$ (denoted as
$\mathbb{C}P^{n-1}$). A point of a $\mathbb{C}P^{n-1}$ is the
equivalence class of vectors in $\mathbb{C}^n$ constructed considering
two vectors for which holds the following relation
\begin{equation}\label{cpd}
(Z_1,Z_2...,Z_n) = (\lambda Z_1,\lambda Z_2...,\lambda Z_n) \;\;\;
\lambda \in \mathbb{C}
\end{equation}
equivalent.
To have an example, the set of pure states of a qubit is the set of points of a
complex projective space of dimension 1 (i.e. $\mathbb{C}P^1$). This manifold is indeed isomorphic to the
ordinary sphere in $\mathbb{R}^3$ and is usually referred to as the
Bloch sphere. Projective spaces over a generic field of numbers
$\mathbb{K}$ can be defined in the same way given above for the complex
case but with the field $\mathbb{K}$ in place of $\mathbb{C}$.
In section
\ref{mqt} we show that if a probabilistic theory satisfies
Distinguishability, Conservation then either is classical or it is a theory in which pure states are points of a
projective space over a generic field of numbers $\mathbb{K}$. Among these
theories we find quantum theory but also quantum theory over real
numbers and quantum theory over quaternions. In section \ref{qt}
it is shown that the only non classical theory in this
class satisfying axioms reversibility and composition to describe composite systems is
quantum theory. The last two
sections are devoted to discussion and conclusion.

\section{Probabilistic Theories}\label{pt}

A generic probabilistc theory is a mathematical
framework in which it is possible to model any experimental
set up and to calculate probabilities for all the possible
configurations of a set up.
In such a framework preparation, transformation and
measurement devices are represented by collections of outcomes, e.g.:
\begin{equation*}
\rho = \{\rho_i\}_{i \in X}\;\; M = \{a_j\}_{j \in Y} \;\;\; T = \{\mathscr{T}_k\}_{k \in Z}
\end{equation*}

where $\rho,M,{T}$ are respectively a preparation,
transformation and measurement while $\phi_i,
\mathscr{T}_k, a_j$ represent the corresponding outcomes in some outcome sets $X,Z,Y$.

An outcome set of a physical device, in general, is not something that have a well
defined probability distribution on its own. The probability
distribution of the outcome set of a
measurement device in an experimental setup,
depends on the settings of the preparation device. This is
clear since if we perform a measurement on a system prepared in some way we have a given probability distribution of the measurement
outcomes, while, if we perform the same measurement on a system prepared in
some different way we obtain, in general, a different probability
distribution. 

Given a preparation and a measurement, $\rho =
\{\phi_i\}_{i \in X}, M = \{a_j\}_{j \in Y}$ we define their
composition as:
\begin{equation} \label{probabilities0}
\circ : (a_j,\rho_i) \rightarrow a_j \circ  \rho_i  \;\; \forall (i,j) \in X
\times Y
\end{equation}  
Since to a preparation followed by a measurement outcome is always
associated the probability of the measurement outcome given that
preparation we can associate to $a_j\circ\rho_i$ the probability of
seeing measurement outcome $a_j$ when performing measurement
$\{a_j\}_{j\in Y}$ on a system prepared in state $\rho_i$, namely:
\begin{equation}\label{probabilities1}
p:  a_j\circ\rho_i \rightarrow p(a_j|\rho_i)
\end{equation}
To use a notation that resembles the braket notation of standard
quantum theory we will also define:
\begin{equation}
(a_j|\rho_i) : = p(a_i|\rho_j)
\end{equation}
Composing the two maps $\circ$ and $p$ we can define a new
map $M$ turning any pair $(a_i\rho_j)$ into a probability:
\begin{equation}\label{probabilities}
M : (a_j,\rho_i) \rightarrow (a_j|\rho_i)
\end{equation}
The probabilistic structure defined by (\ref{probabilities}) turns
every $\rho_i$ into a function from measurement outcomes to probabilities, given
by $ M[(\cdot,\rho_i)]$.
If $\rho_i, \rho_i' $ induce 
the same function, then it is impossible to distinguish between them
from the statistics of an experiment. This
means that the two outcomes of the preparation device are equivalent: accordingly,
we will take equivalence classes with respect to this equivalence relation. We will thus identify the outcomes
$\rho_i,\rho_i' $ with the corresponding function $\rho_i$
and will call it \emph{state}.  Accordingly, we will refer to
preparation devices as collections of states $\{\rho_i\}_{i \in \rX}$.
The same construction holds for measurements: every measurement outcome $a_j $ induces a
function from preparations to probabilities, given
by $M [(a_j,\cdot)]$.
If two outomes $a_j, a_j'$ induce the same function, then it is impossible to
distinguish between them from the statistics of an experiment. This means
that the two outcomes are equivalent: accordingly, we will take equivalence classes
with respect to this equivalence relation. We will thus identify the outcome $a_j$ with the corresponding
function and we will call it \emph{effect}.  Accordingly, we will refer to
measurement devices as collection of effects $\{a_j\}_{i \in \rY}$. 

The state of a system provides the information regarding the probabilities of all
the possible outcomes in all the possible measurements that can be
performed on a system prepared in a given configuration. Thus the
state of a system can be represented by a list of probabilities that,
in principle, could contain an infinite number of entries. Hence for a state $\rho$ of a
system we can write:

\begin{equation}\label{sta}
 \rho = \begin{pmatrix}
\vdots       \\
p_{a_j}       \\
\vdots  
\end{pmatrix} 
\end{equation}
where $p_{a_j}$ is the probability of the effect labeled by $a_j$ given the state $\rho$. $\rho$ is thus represented by the
vector (\ref{sta}) and two different state vectors of a given system will
differ at least for one of their entries. In principle the number of entries for a state
expressed as in (\ref{sta}) is infinite since,
typically, there is a
continuum of possible measurements associated to a system. For example the spin of a particle
can be measured in all directions in space and thus we have a continuum of possible
measurements for the spin. However in quantum theory a restricted subset of
entries suffices to specify a state as in (\ref{sta}). For a qubit, the
number of such entries is 4, i.e. the number of real parameters that
are necessary in order to specify a hermitian matrix acting on
$\mathbb{C}^2$. In \cite{hardy3} it is argued this kind of
compression is a general feature of all physical theories and it is
taken as a starting point to formulate a framework for quantum
gravity. In this paper we will assume that the number of entries in
the list of probabilities representing a state of a physical system as
in (\ref{sta}) is finite.
 
From the above considerations we have
that states can be represented as elements of a real finite
dimensional vector
space while effects can be represented in the vector space
constituting the dual of the state space. As a consequence
transformations of a system A can be represented as operators defined on the vector
space in which are represented the states of A.
\begin{definition}\label{dimen}
The dimension of the real vector space where states of
a system are represented as vectors of probabilities is \emph{the
dimension of the system}.
\end{definition}
For a quantum system seen as an object of a probabilistic theory the dimension of the hilbert
space associated to a system does not coincide with the dimension of
the system. The dimension of a quantum system in this framework is the dimension of the real vector space in
which are defined density matrices representing states. This is the real vector space generated by hermitian operators
acting on the hilbert space associated to the system. The dimension
of such real vector space is $d = n^2$ where $\mathbb{C}^n$ is the
hilbert space associated to the system. This notion of dimension
of a system is introduced in \cite{axioh}. 

The probability distribution of the effects $A = \{a_j\}_j$ in measurement
$A$ always depends on
what state is prepared. It then must hold that the
probability of happening of state $\rho_i$ must be independent of the measurement
that is performed. If it were not so then the probability distribution of the effects in
a measurement would depend on the state of the phsyical system
and, at the same time, would be something on which the state of
the system depends. If this was the case then states and effects would
be related in a non linear way and the function associating pairs of
state and effect to a probability would be non linear. This would
prevent the possibility to form mixture of states (see below) and of representing
states in a real vector space. Since the
probability of happening of state $\rho_i$ must be independent of the
measurement performed we must have:
\begin{equation}\label{causa}
(\sum_{j\in Y} a_j|\rho_i) = (\sum_{l\in S} b_l|\rho_i) \;\;\forall\;\;
\{a_j\}_{j\in Y},\{b_l\}_{l\in S}
\end{equation}
where $\sum_{j\in Y} a_j,\sum_{l\in S} b_l$ represents respectively
a singleton (i.e. a single outcome measurement) constituted by the coarse graining of all the outcomes in measurements $A = \{a_j\}_{j\in Y}$ and $B = \{b_l\}_{l\in
  S}$. Condition (\ref{causa}) is
stated in \cite{puri} and is associated to \emph{causality}.  We now give
the following: 
\begin{definition}
Given any measurement $M = \{m_j\}_{j\in Y}$, the
coarse graining of the effects in $M$, $\cup_{j\in Y} \; m_j$, is called deterministic effect. 
\end{definition} 
We now state the following characterization of (\ref{causa}) 
\begin{proposition}
(\ref{causa}) is equivalent to require that the deterministic effect is
unique for all measurements.
\end{proposition}
This proposition is proved in \cite{puri}.

Given (\ref{causa}) we must have that the probability distribution of the states of a
preparation device $\rho = \{\rho_i\}_{i\in X}$ must be independent of
the settings and outcomes on other devices in any experiment. This
implies that to every preparation $\rho = \{\rho_i\}_{i\in X}$ must be
associated a probability distribution $\{p_i\}_{i \in X}$ for the
corresponding states. Hence every state $\rho$ can be
represented as a convex combination of other states, namely:
\begin{equation}\label{conv}
\rho = \sum_{i\in X} p_i \rho_i
\end{equation}
From (\ref{conv}) and the fact that states can be represented
as in (\ref{sta}) we have that the state space of any physical system
is a compact convex set. 
This leads us immediately to the following definition:
\begin{definition}
A state is \emph{mixed} if it can be represented as a convex
combination of at least two other states. A state is \emph{pure} if
the convex combination representing it contains only one state.
\end{definition}
Since we have that $\mathscr{T}_k \rho = \sum_{i\in X} p_i
\mathscr{T}_k \rho_i$ for every transformation $\mathscr{T}_k$ and
every state $\rho$ we must have that transformations are linear
operators.
Note however that the convex decomposition of a state $\rho$ is not
required to be unique. There can be many ways to express the same
state $\rho$ as a convex combination of other states.

\begin{definition}\label{refine}
The \emph{refinement of state $\rho$} is defined to be
the set of states that can appear in a convex decomposition
representing $\rho$.  
\end{definition}

We will denote the refinement set of a state $\rho$ as $F_{\rho}$.
The notion of refinement of a state will play a crucial role in this
derivation. It is introduced in the framework of probabilistic theories
in \cite{axioq}. From the mathematical point of view, the refinement
of a state coincides with the notion of \emph{face} of a convex
set. A face $F$ of a convex set, $S$, is defined to be a convex subset
of $S$ such that given two points, $x_1,x_2 \in S$ if $\lambda x_1 +
(1-\lambda) x_2 \in F$ then $x_1,x_2 \in F$. From definition
\ref{refine}, we see that the refinement of a state $\rho$ is a face
of the (convex) state space of the system of which $\rho$ represents a
preparation.

\begin{definition}\label{complmix}
A state $\rho$ is \emph{completely mixed} relatively to a set of states $S$,
if all the states in $S$ can appear in the convex decomposition of $\rho$.    
\end{definition}

A physical system can be used to store
information. Storage of information into a physical system is possible
if it can be defined a protocol that can be used to read that
information. To define such a protocol we give the following:  

\begin{definition}\label{perfdist}
The set of states $\{\rho_i\}_{i = 1}^N$ is \emph{perfectly distinguishable}
if there exists a measurement $A =  \{a_j\}_{j = 1}^N$ such that
$(a_j|\rho_i) = \delta_{ij}$.
The set of effects $\{a_j\}_{j = 1}^N$ is a set of \emph{perfectly discriminating effects}.
\end{definition}

Given the above definition we see that if the set of states of a system contains at least two
perfectly distinguishable states, the system can be used to store
information. Referring to the above definition, if one prepares a
state belonging to a set of perfectly distinguishable
states, $\{\rho_i\}_{i = 1}^N$, say $\rho_{i_0}$, then performing the measurement
$A = \{a_j\}_{j = 1}^N$ one can say with certainty that the state
prepared was $\rho_{i_0}$ upon seeing the effect $a_{i_0}$. This in turn
defines a protocol to read the information that can be stored in the
physical system. A
list of $L$ symbols choosen among the set $\{i\}_{i=1}^N$ can be
stored in $L$ copies of the system with preparations choosen in the set
of states $\{\rho_i\}_{i=1}^N$.

In Quantum theory a system having states defined on a Hilbert space of
dimension $n$, $\mathbb{C}^n$ has $n$ perfectly distinguishable
states. 

We conclude this section 
with
the following two definitions 

\begin{definition}
The set of perfectly distinguishable states $\{\rho_i\}_{i = 1}^N \in
S$, where $S$ is a generic set of states, is
\emph{maximal} in $S$ if there does not exist a state $\sigma$ in $S$
such that $\{\rho_i\}_{i = 1}^N \cup \sigma$ is a perfectly
distinguishable set of states.
\end{definition}

\begin{definition}
The \emph{information capacity} of a set of states is the
cardinality of the largest maximal set of perfectly distinguishable states in
that set. 
\end{definition}

\section{Axioms for quantum theory}\label{aqt}

In what
follows we will impose a set of four axioms that are very natural for a
generic probabilistic theory in which systems can be used to store
information. It will be shown in the following sections that the
only non classical
theory satisfying all these axioms is quantum theory. 
The axioms are called Distinguishability,
Conservation, Reversibility, Composition. The starting point to
formulate them is that every physical system can be used to
store information. To understand the meaning of this consider a system
as simple as a die. On every face of a
die there is a certain number of dots from one to six. Suppose that
one has to store and retrieve a text written using an
alphabet of six elements $\{a,b,..,f\}$. He can encode every letter of the
alphabet into a number from one to six. The text is an ensemble of
letters that have a certain probability of appearing
$\{p_a,p_b,..,p_f\}$. This text can thus be stored into an ensemble of
dice that are prepared according to the probability distribution
$\{p_a,p_b,..,p_f\}$. The preparation of the ensemble for storage of the text can be performed leaning
each die in the ensemble on a surface in such a way that the face
corresponding to the stored letter is hided. One can
retriveve the text simply looking at which is the hided face for every die. 
The above protocol to store a text into an ensemble of
systems can be accomplished with a quantum system with
hilbert space dimension six as well. One can choose a basis
for the system, $\{|i\rangle\}_{i=1}^6$ and encode every letter of the
alphabet into one of these states. Storage of the the text into an
ensemble of physical systems can be performed preparing each quantum system in the ensemble in one of these
states. The text can thus be retrieved
measuring each element of the ensemble of quantum systems in the above
basis.

We are now going to state and explain the axioms.

\begin{quote}\label{dist} {\bf Axiom 1 - Distinguishability} Every
  state that is not completely mixed relatively to a given refinement
  set is perfectly distinguishable from another state in that set. 
\end{quote}

This axiom is a stronger version of axiom distinguishability used in
\cite{axioq}. 
Recall that the notion
of refinement set of a state coincides with the notion of face of a
convex set. This axiom
tells that if $\rho$ is not completely mixed in a given set of states
of a system $F$ constituting a face,
then there exist another state $\phi$ in that face and
a measurement $A = \{a_{\rho}, a_{\phi}\}$ that can be performed on
system $\mathcal{S}$ such that
$(a_{\rho,\phi}|\rho,\phi)=\delta_{\rho,\phi}$. To see this in a simple context consider again
a die. Take a state of a die described by the statistical
mixture of the face "one" (here we mean the physical face of the die
that has one dot on it not the
mathematical concept of convex analysis) and the face "six" with
probabilities $\{p_{\text{one}},
p_{\text{six}}=(1-p_{\text{one}})\}$. This state is not completely
mixed since there exist faces of the die that are not used in the
above mixture, e.g. the face with three dots that we denote face "three". The axiom tells that face "three" can be
used with the above mixture, $\{p_{\text{one}},
p_{\text{six}}=(1-p_{\text{one}})\}$, to store one bit of
information. Indeed one can choose to assign the logical value "0" to
both faces "one" and "six" and the logical value "1" to face "three". 
This axiom holds classically. Recall that the set of states of a system with
information capacity $n$ is a simplex in $\mathbb{R}^{n-1}$.
The refinement of a mixed state $\rho$ that is not completely mixed, constitutes a simplex in
$\mathbb{R}^{m}$ with $m<n-1$. Any pure state $\phi$ not belonging to
$F_{\rho}$ is represented by a vector in $\mathbb{R}^{n-1}$
orthogonal to the subspace in which is embedded $F_{\rho}$. Hence we
have that $\phi$ is perfectly distinguishable from all states in
$F_{\rho}$ hence also from $\rho$ itself. The axiom
holds also in quantum theory. The refinement set of a mixed
state $\rho$ of a system with information capacity $n$, is the convex
hull of a $\mathbb{C}P^{m-1}$ where $F_{\rho}$ has information
capacity $m \leq n$. Any pure state $\phi$ not belonging to $F_{\rho}$ is
represented by a complex vector orthogonal to the subspace
representing $F_{\rho}$. Hence $\phi$ is perfectly distinguishable
from all the pure states in $F_{\rho}$ and thus from $\rho$ itself.





\begin{quote}\label{alph} {\bf Axiom 2 - Conservation} The information capacity of the refinement
  of any mixture of two states is less than or equal to the sum of the
  information capacities of the refinements of the two states
  composing the mixture, with equality if the states
  are perfectly distinguishable.
\end{quote}

Let
$\eta = p \sigma + (1-p) \rho$. The axiom tells that
there cannot exist sets of states that are not perfectly
distinguishable in $F_{\sigma}$ and sets of states
that are not perfectly distinguishable in $F_{\rho}$ that become
sets of perfectly distinguishable states if regarded as states of
$F_{\eta}$. Moreover if $\rho$ and $\sigma$ are
perfectly distinguishable states, then the information capacity of $F_{\eta}$
is the sum of the information capacities of $F_{\rho}$ and
$F_{\sigma}$. Hence the number of pure states usable to store
information in $F_{\eta}$ is the sum of the number of pure states
usable to store information in $F_{\sigma}$ and the number of
pure states usable to store information in $F_{\rho}$. To have an
example consider two mixtures of different faces of a die, say a mixture of
"one" and "six" $\{p_{\text{one}},
p_{\text{six}}=(1-p_{\text{one}})\}$ and a mixture of "two" and
"four" $\{p_{\text{two}},
p_{\text{four}}=(1-p_{\text{two}})\}$. The axiom states that if we prepare
the following mixture $\{qp_{\text{one}},qp_{\text{six}},tp_{\text{two}},tp_{\text{four}}\}$
with $q = 1- t$,
then the number of values that can be used to store
information in this mixture is not greater than four that is the
number of values used in
preparing the above two mixtures, namely, "one","six","two","four".  
This axiom clearly holds in classical theory. $\sigma$ and $\rho$ are such
that $F_{\sigma}$ and $F_{\rho}$ are two simplexes with $s$ and $r$
verteces respectively and live in $\mathbb{R}^{s-1}$ and
$\mathbb{R}^{r-1}$. The refinement set of $\eta = p \rho +
(1-p)\sigma$ has dimension $s+r-t-1$ where we assumed the intersection of
$F_{\rho}$ and $F_{\sigma}$ to be a simplex in
$\mathbb{R}^{t-1}$.
$F_{\eta}$ is a simplex in $\mathbb{R}^{s+r-t-1}$ and has information capacity $r+s-t$ where
$t\geq 0$ with equality iff $\rho$ and $\sigma$ are perfectly distinguishable. The axiom holds in quantum theory as well. Take a convex
combination of two density matrices $\rho$ and $\sigma$. $F_{\rho}$
and $F_{\sigma}$ are the convex hull of the space of rays in
$\mathbb{C}^r$ and $\mathbb{C}^s$ respectively. $F_{\eta}$ is the
convex hull of the rays belonging to the smallest complex vector space containing both
$\mathbb{C}^r$ and $\mathbb{C}^s$, i.e. $\mathbb{C}^{r+s-t}$ where
$\mathbb{C}^t$ is the intersection of $\mathbb{C}^r$ and
$\mathbb{C}^s$. The information capacity of $F_{\eta}$ is thus $r+s-t$
with $t \geq 0$ with equality iff $\rho$ and $\sigma$ perfectly distinguishable.


\begin{quote}
{\bf Axiom 3 - Reversibility} For any two pure states
of a system, $\phi, \psi$, there exists a reversible transformation
$\mathscr{D}$ such that $\phi = \mathscr{D} \psi$.
\end{quote}

The significance of the above axiom is simply that one can
transform any pure state into any other with a reversible
transformation. This axiom clearly holds classically since one can
transform any pure state of a simplex in $\mathbb{R}^{n-1}$ into any other applying a
transformation in the symmetric group $S_n$ (i.e. the group whose
elements are the permutations of $n$ objects  with composition rule
being sequential composition of permutations). This axiom also holds in quantum
theory since the set of pure states of a system with hilbert space
$\mathbb{C}^n$ is such that any pure state can be transformed into any
other with a transformation belonging to the group SU$(n)$.


\begin{quote}
{\bf Axiom 4 - Composition} If $d_A$ and $c_A$ are the dimension
and the information capacity of system
$A$ then
\begin{equation*}
d_{ABC..} = d_Ad_Bd_C\cdots 
\end{equation*}
and
\begin{equation*}
c_{ABC..} = c_Ac_Bc_C\cdots
\end{equation*}
where $ABC..$ is the system composed of systems $A,B,C..$ 
\end{quote}
The first part of the above axiom is called local tomography.
Every state of a system in a probabilistic theory can be represented
as a list of a certain number of probabilities obtained performing an equal number of different
measurements on the system in that state (see (\ref{sta})). The number of measurements that are
sufficient in order to uniquely determine a state as a vector of
probabilities is the number of degrees of freedom of the
system that is equal to its dimension by definition \ref{dimen} (see
also \cite{axioh}). The first part of axiom composition requires that the number of
degrees of freedom of a composite system be the product of the number
of degrees of freedom of the components. An important consequence of
this axiom is that the real vector space in which it
can be represented the state space of a composite system $\mathcal{S}_{AB}$ is the vector space tensor product of the vector
spaces in which are represented the state spaces of the component
systems, $\mathcal{S}_A$, $\mathcal{S}_B$. The second part states
simply that for a system composed of a certain number of component systems,
the information capacity of the composite system is the product of the
information capacities of the components. To see that the above axiom
holds classically note that the state space of a composite system is a
simplex in the real vector space
tensor product of the vector spaces where live the components. Hence
the first part of the axiom holds. Since
the information capacity of a classical system is equal to the number
of degrees of freedom of the system we have that also the second part holds
classically. The axiom holds in quantum theory as well. Indeed a density matrix of a composite system $\rho_{AB}$ is
a hermitian operator defined on the space of hermitian operators tensor product of the spaces where
live density matrices for the component systems $A$, $B$. These
latter spaces are those where live
operators acting on $\mathbb{C}^{n_A}$ and $\mathbb{C}^{n_B}$
respectively. Hence the
first part of the axiom holds. The second part holds as well since the
information capacity of the state space of a quantum system
coincides with the dimension of the complex vector space on which are defined
density matrices describing states of the systems.
$\mathbb{C}^{n_A}$ and $\mathbb{C}^{n_B}$ are hilbert spaces of
component systems with information capacity $n_A$ and $n_B$
respectively, and 
$\mathbb{C}^{n_A} \otimes \mathbb{C}^{n_B}$ is the hilbert space for
the composite system with information capacity $n_An_B$.



\section{Proof of the main result}

\subsection{Many quantum theories}\label{mqt}


In this section we show that a probabilistic theory satisfying axioms
1,2 either is classical, or is a probabilistic theory in which pure states
are points of a projective space. The fact that classical probability
theory satisfyies axioms 1,2 is already proved in the previous section. We thus are going to prove the remaining alternative, namely,
pure states of a system with infromation capacity $n+1$ of a probabilistic theory satisfying axioms
1,2 are points in a projective space of dimension $n$. From this fact
it will follow that this class of theories is such that pure states are elements in a vector space defined over a
generic field of numbers (more precisely, a generic field of numbers is called \emph{division
ring} ). This implies that for theories in this class it holds
the superposition principle. Pure states of such
theories can be represented by elements of a vector space in which all states
in a subspace $A$ are perfectly distinguishable from all states in a
subspace $B$ disjoint from $A$ and linear combinations of elements in
disjoint subspaces represent allowed pure states.

A projective space of generic dimension $n$ is defined in the following \cite{casse}:

\begin{definition}\label{proj}
{\bf Projective space of dimension n}

An arbitrary set $\mathcal{S}$, together with a family of subsets, $F^j$, that are called subspaces of dimension $j$, is a projective space of dimension $n$
if the following holds:

\emph{(i)} The only $-1$-dimensional subspace is the empty set.

\emph{(ii)} 0-dimensional subspaces of $F^n$ are the point subsets of $\mathcal{S}$.

\emph{(iii)} There is a unique subspace of dimension $n$, $F^n$. 

\emph{(iv)} If $F^r$ and $F^s$ are two subspaces of $F^n$, and $F^r$ is
  contained in  $F^s$ then $F^r\equiv F^s$ if and only if $r=s$.        

\emph{(v)} Given two subspaces $F^r$ and $F^s$ of $F^n$, if $F^t$ is the intersection of $F^r$ and $F^s$ then $F^t$ is a subspace of $F^n$. 

\emph{(vi)} Given two subspaces $F^{r}$ and $F^{s}$ of $F^{n}$, if
  $F^m\equiv F^{r} \cup F^s$ and $F^t$ is the intersection of $F^r$ and $F^s$, then: $s+r=t+m$.

\end{definition}

In the following we will
show that for pure
states and for a family of
subsets of these states of a system with information capacity $n+1$ described by a
probabilistic theory satisfying axioms 1,2, definition \ref{proj}
holds. 
To prove that definition \ref{proj} holds for pure states of a system
of a probabilistic theory we have to define the notion of subspace in
this context. In what follows we will denote $\mathcal{S}$ the set of
states of a generic system.

\begin{definition}\label{subspa}
Given the set of states of a generic system $\mathcal{S}$ with information capacity $n+1$ and any state $\rho \in \mathcal{S}$ such that
the cardinality of the largest maximal set of perfectly distinguishable states
in $F_{\rho}$ is $j+1\leq n+1$, the set $F_{\rho}$ is called \emph{subspace} of $\mathcal{S}$ with
dimension $j$ and
denoted $F_{\rho}^j$  
\end{definition}

\begin{definition}\label{empty}
The empty set $\emptyset$ is defined to be a subspace of the set of
states of a system $\mathcal{S}$ with dimension -1 and is denoted $F^{-1}$.
\end{definition}

\begin{definition}\label{point}
Pure states in $\mathcal{S}$ are defined to be subspaces of
dimension $0$ and denoted $F^0$.
\end{definition}

In the following we will suppose that $\mathcal{S}$ is the set of
states of a generic system with information capacity $n+1$ of a generic probabilistic
theory satisfying axioms 1,2. Lemmas
\ref{3}-\ref{5} and theorem \ref{6} are needed to prove
theorem \ref{theoproj}, containing the most important part of this section.

\begin{lemma}\label{3}
The only subspace of dimension $n$ in $\mathcal{S}$ is $\mathcal{S}$ itself. 
\end{lemma}
\emph{Proof:} Let $F_{\theta}^n$ be a subspace of dimension $n$ in
$\mathcal{S}$ generated by the refinement of a mixed state $\theta$.
We now prove the thesis showing that $\theta$ is completely mixed
in $\mathcal{S}$. To see this suppose it is not so. Then $\theta$ is
not completely mixed.
From axiom distinguishability $\theta$ is perfectly
distinguishable from some state $\phi$. The state $p\theta +
(1-p) \phi$ is in $\mathcal{S}$ and its refinement has information
capacity greater than or equal to $n+2$ from axiom conservation.
This contradicts the hypothesis that
$\mathcal{S}$ has information capacity $n+1$ and proves the thesis.
$\blacksquare$

\begin{lemma}\label{4}
If $F_{\rho}^r$ and $F_{\sigma}^s$ are both subspaces of $\mathcal{S}$ and $F_{\rho}^r
\subseteq F_{\sigma}^s$ then $r \leq s$. $r=s$ iff $F_{\rho}^r
= F_{\sigma}^s$
\end{lemma}
\emph{Proof:} If $F_{\rho} \subseteq F_{\sigma}$ then the information
capacity of $F_{\rho}$ cannot exceed that of $F_{\sigma}$. If $F_{\rho}^r$ and
$F_{\sigma}^s$ are the same subspace then clearly
$r=s$.  To prove the converse note that by hypothesis we have that for all
$\phi \in F_{\rho}^r$ we must have $\phi \in F_{\sigma}^s$. Now
suppose that $r=s$ and $F_{\rho}^r \subset F_{\sigma}^s$. This means
that $\rho$ is not completely in $F_{\sigma}$. From
axiom distinguishability $\rho$ is perfectly distinsguishable from
some state $\phi$ in $F_{\sigma}^s$. The refinement of the state $\omega = p \rho
+ (1-p) \phi$ must have information capacity equal to $s+2$ by hypothesis and
from axiom conservation. But this is absurd since $\omega \in
F_{\sigma}^s$ and $F_{\sigma}^s$ must have information capacity $s+1$
by definition \ref{subspa}. Hence if $F_{\rho}^r
\subseteq F_{\sigma}^s$ and $r=s$ then we must have $F_{\rho}^r =  F_{\sigma}^s$.
$\blacksquare$

\begin{lemma}\label{spanbit}
The refinement of a mixture of any two pure states of a system
has information capacity two. 
\end{lemma}
\emph{Proof:} From axiom conservation the refinement of a convex
combination of any two pure states of a system cannot have information
capacity greater than two. Since any of the two pure states is not
completely mixed by definition, from axiom distinguishability we have
the thesis.
$\blacksquare$

\begin{definition}\label{inters}
Given two subspaces of a system $\mathcal{S}$, $F_{\rho}^r$ and $F_{\sigma}^s$, we denote their
\emph{intersection} as
$F_{\rho}^r \wedge F_{\sigma}^s$
\end{definition}

\begin{lemma}\label{5}
Given two subspaces of $\mathcal{S}$, $F_{\rho}^r$ and $F_{\sigma}^s$,
$F_{\rho}^r \wedge F_{\sigma}^s$
is a subspace of $\mathcal{S}$.
\end{lemma}
\emph{Proof:} The intersection of two convex sets is a convex
set. Hence there exists a state $\tau$ that is completely mixed in the
set $F_{\tau} = F_{\rho} \wedge F_{\sigma}$. The set of states of $\mathcal{S}$
in $F_{\tau}$ constitutes a subspace of
$\mathcal{S}$ with information capacity greater than or equal to two
from axiom distinguishability.
$\blacksquare$

\begin{definition}\label{span}
Given two subspaces, $F_{\rho}^r$ and $F_{\sigma}^s$ we define their
\emph{span} and denote it as
$F_{\rho}^r\vee F_{\sigma}^s$ the set of states in the refinement of a convex
combination of $\rho$ and $\sigma$.
\end{definition}

\begin{theorem}\label{6}
Given two distinct subspaces of $\mathcal{S}$, $F_{\rho}^r$ and $F_{\sigma}^s$,
if $F_{\eta}^m = F_{\rho}^r \vee F_{\sigma}^s$ and $F_{\tau}^t
= F_{\rho}^r \wedge F_{\sigma}^s$ with $\eta = p \sigma + (1-p)
\rho$ and $\tau$ completely mixed state in $F_{\rho} \wedge F_{\sigma}$, then $r + s = t + m$
\end{theorem}
\emph{Proof:} In the case both $\rho$ and $\sigma$ are pure states the
thesis holds from lemma \ref{spanbit} and definition
\ref{empty}. 

Suppose that only one of them, say $\sigma$, is
mixed. Then either $\rho$ is in $F_{\sigma}$ and then $\tau=\rho$ or
$\rho$ is not in $F_{\sigma}$ and $\tau$ is the empty set. In both
these cases the thesis trivially holds.

Suppose now that both $\rho$ and $\sigma$ are mixed states.
By hypothesis and axiom distinguishability $\tau$
is perfectly distinguishable from some state $\phi_1$  in $F_{\rho}^r$ that we choose
w.l.o.g. pure. Let
$\tau_1 = p \tau + (1-p) \phi_1$, $0<p<1$. From axiom
conservation the information capacity of $F_{\tau_1}$ is $t+2$. Thus
we have constructed the subspace $F_{\tau_1}^{t+1}$. The state
$\sigma_1 = p \sigma + (1-p) \phi_1$, $0<p<1$, is such that
$F_{\sigma_1}$ has information capacity less than or equal to $s+1$
from axiom conservation. Moreover, from axiom distiguishability, there
exists a pure state $\psi_1\in F_{\sigma_1}$ that is perfectly
distinsguishable from $\sigma$ and such that the
state $\sigma_1' = p \sigma + (1-p) \psi_1$ has refinement
$F_{\sigma_1'}$ with information capacity equal to $s+1$. Since by
hypothesis $\sigma_1' \in F_{\sigma_1}$ we must have that the
information capacity of $F_{\sigma_1}$ is $s+1$. In this way we have
constructed the subspace $F_{\sigma_1}^{s+1}$. 
Now we can have either $\tau_1$ completely mixed in $F_{\rho}^r$ or not. If
the latter is the case then, in the same way we constructed
$F_{\tau_1}^{t+1}$ with $F_{\tau}^t$ and $\phi_1$ we construct
$F_{\tau_2}^{t+2}$ with $F_{\tau_1}^{t+1}$ and a pure state $\phi_2$
in $F_{\rho}^r$. In the same
way as before we can also construct $F_{\sigma_2}^{s+2}$ with
$F_{\sigma_1}^{s+1}$ and $\phi_2$. Iterating this procedure we will
end for some finite $k$ to a state $\tau_k = p\tau_{k-1} + (1-p)
\phi_k$, $0<p<1$ that is completely mixed in $F_{\rho}^r$. This happens when
there are no more states in $F_{\rho}^r$ outside $F_{\tau_k}^{t+k}$. $k$ is finite since, if it were not so, we
would have infinite perfectly distinguishable states in
$F_{\rho}^r$ contradicting the hypothesis ($r$ is finite since $n$
is). Iterating the above construction we also
obtain the subspace $F_{\sigma_k}^{s+k}$ where $\sigma_{k} = p
\sigma_{k-1} + (1-p) \phi_k$, $0<p<1$. By construction we have
$F_{\tau_k} \subset F_{\sigma_k}$. Moreover $F_{\sigma_k}^{s+k}$ contains every
state in $F_{\sigma}^s$ and there cannot exist states in $F_{\rho}^r$
outside $F_{\sigma_k}^{s+k}$ since this would contradict the
facts that $F_{\tau_k}^{t+k} = F_{\rho}^r \subset F_{\sigma_k}^{s+k}$. But this in turn means that every
state in $F_{\eta}^m$ must pertain to $F_{\sigma_k}^{s+k}$ and thus $\sigma_k$
is completely mixed in $F_{\eta}$. All this implies that $F_{\tau_k}^{t+k} =
F_{\rho}^r$ and $F_{\sigma_k}^{s+k} = F_{\eta}^m$ thus $k=r-t$ and $m=s+k$. Thus we find $m = s+r-t$ and prove the
thesis.
$\blacksquare$

\begin{theorem}\label{theoproj}
If the states of a system with information capacity $n+1$ are
described by a probabilistic theory satisfying axioms 1,2 then pure
states of that system are points of a projective space of
dimension $n$. 
\end{theorem}
\emph{Proof:} For a system $\mathcal{S}$ with
information capacity $n+1$ satisfying axioms 1,2 there
exists a family of subspaces $F^j$ of dimension $j$ such that:

\emph{(1)-(2)} (i)-(ii) in definition \ref{proj} hold by definitions
\ref{empty}, \ref{point}.

\emph{(3)} (iii) in definition \ref{proj} holds by lemma \ref{3}

\emph{(4)} (iv) in definition \ref{proj} holds by lemma \ref{4}

\emph{(5)} (v) in definition \ref{proj} holds by lemma \ref{5}

\emph{(6)} (vi) in definition \ref{proj} holds by theorem \ref{6}
$\blacksquare$

Projective spaces can be very wild geometrical objects. All spaces
in which points are one dimensional subspaces of a vector
space defined over some field of numbers constitute projective spaces
but not all projective spaces can be defined in this way. Examples
of spaces in these class are the so called non-Desarguesian planes
\cite{casse,w10,w17}. Fortunately, the
following theorem characterizes projective spaces of dimension greater
than two, and, as a
consequence, the set of pure states of systems described by theories
obeying axioms 1,2 with information capacity greater than three \cite{w08,w10,w17}.  

\begin{theorem}\label{veblen}
\emph{Veblen and Young theorem}

If the dimension of a projective space is $n \geq 3$ then it is
isomorphic to a $\mathbb{K}P^n$, i.e. a projective space of dimension $n$ over some division ring $\mathbb{K}$.
\end{theorem} 
The proof of this mathematical result is not in the scope of this paper.
A projective space over a division ring $\mathbb{K}$ is defined as a space
whose points are one dimensional subspaces of a vector space defined
over the division ring $\mathbb{K}$. A division ring (also called skew
filed) is a mathematical
generalization of the concept of field of numbers (e.g. reals or complex numbers),
in which multiplication is not needed to be commutative. The most popular example of skew field in which multiplication is not
commutative are the quaternions. In the rest
of the paper we will refer to a skew filed of numbers simply as a
``field'' for brevity.

Theorem \ref{veblen} characterizes the state space of systems with
information capacity greater than or equal to four described by a generic
probabilistic theory satisfying axioms 1,2.
We have thus proved that in a theory satisfying axioms 1,2, systems
with information capacity greater than three either are classical or are
such that pure states information capacity $n+1$, $n>2$, are
points of a $\mathbb{K}P^n$. These can be seen as elements of a vector space defined over a
generic skew field $\mathbb{K}$. In this landscape a set of pure
perfectly distinguishable states $\{\phi_0,\phi_1,...,\phi_n\}$ of a
system with information capacity $n+1$ is a set of $n+1$ disjoint
subspaces of the vector space in which are represented pure states of
the system. If a system can be
in one of the states $\{\phi_0,\phi_1,...,\phi_n\}$ then it can also
be found in a superposition of these states $\phi_s = \sum_{i=0}^nk_i\phi_i$
with $\{k_{i}\} \in \mathbb{K}$ since this linear combination
represents an element of the vector space and thus an allowed pure
state. Hence in a probabilistic theory
satisfying axioms 1,2, systems with information capacity greater than
three are either classical (since classical theory
satisfies them) or a generalization of quantum systems in which
superposition principle holds with coefficients not necessarily
complex but belonging to a generic (skew) field of numbers $\mathbb{K}$.
In section \ref{qt} it will be proved that this characterization also
holds for system with information capacity three or two.

\subsection{State space of an elementary system}\label{elsys}

An elementary system is a system with information capacity two. We
will denote the set of states of an elementary system $\mathcal{S}_2$.
We now show that all theories satisfying axioms 1-3 are
such that the set of pure states in the refinement
of a state with information capacity two, constitutes a sphere
embedded in euclidean space of some dimension $d_2$. If $\rho$ is a
completely mixed state in $\mathcal{S}_2$ this result implies that the number of
entries of a real vector
representing a
normalized state of an elementary system is $d_2+1$ where the
normalization degree of freedom is explicitly taken into account. In quantum theory $d_2=3$.


We are now going to show that, given a state $\rho$, if $F_{\rho}$ has
information capacity two, then states in $F_{\rho}$ constitute a
sphere.

\begin{lemma}\label{ball}
Given a state $\rho$, if $F_{\rho}$ has information capacity two, then
the set of pure states in $F_{\rho}$ is a
sphere.
\end{lemma}
\emph{Proof:}
From axiom reversibility, for any two pure states in $F_{\rho}$,
$\psi$, $\phi$, there exists a transformation $G$ such that $\phi =
G\psi$. The set of transformations mapping pure states in $F_{\rho}$
into pure states in $F_{\rho}$ forms a group. This group must be
compact since vectors representing pure states in $F_{\rho}$
constitute a real representation of it and have bounded entries.
From this fact and the fact that any compact group admits a
representation by means of orthogonal $d\times d$ matrices, we
know that we can represent pure states in $F_{\rho}$ as points of a
$d$-dimensional sphere. We now prove that every point of
the sphere represents a pure state. In order to prove it suppose it is not
so. Then it exists a mixed state, $\sigma$, on the border of the
convex set $F_{\rho}$ that
is not completely mixed. From axiom distinguishability, $\sigma$ must be perfectly
distinguishable from some other pure state, $\phi$, not in $F_{\sigma}$.
$F_{\sigma}$ contains at least two perfectly distinguishable states from
axiom distinguishability since $\sigma$ is by hypothesis mixed, and
there must be a pure state in $F_{\sigma}$ that is perfectly
distinguishable from some other state in $F_{\sigma}$. The state $\omega = p\sigma + (1-p)\phi$, $0<p<1$ is
in $F_{\rho}$ thus by hypothesis $F_{\omega}$ cannot have information
capacity larger than two. Indeed from axiom conservation,
$F_{\omega}$ has information capacity equal to the sum of one plus the
information capacity of $F_{\sigma}$ thus having 
information capacity greater than two. Hence we find a contradiction
proving that the set of pure states in $F_{\rho}$ constitutes a sphere.  
$\blacksquare$

The above lemma permits to conclude that an elementary system
satisfying axioms 1-3 is either classical or is a generalized Bloch
sphere in dimension $d_2$. In the following lemma we will explicit the Bloch representation for
the generalized elementary system. 
\begin{lemma}\label{repbit}
A point $\psi$ of the sphere constituting the
state space of an elementary system has the following form:
\begin{equation*}
\psi= \begin{pmatrix}
2p(x_1) - 1 \\     
2p(x_2) -1 \\
\vdots  \\ 
1
\end{pmatrix} 
\end{equation*}
where $\{x_i\}_{i=1}^{d_2}$ is a set of fiducial effects for $\mathcal{S}_2$.
\end{lemma}
\emph{Proof:} In the representation
where $\psi$ is a vector on the unit sphere in $\mathbb{R}^{d_2}$,
the probability of seeing an effect $\phi$ in a measurement given that it
is prepared state $\psi \in \mathcal{S}_2$ is:
\begin{equation}
E_{\phi}(\psi) = 1/2(1+\phi^T\psi)
\end{equation}
with $\phi$ representing some other unit vector on the sphere.
The orthonormal basis for $\mathbb{R} ^{d_2}$ is:
\begin{equation}\label{orthon}
x_1 = \left[ \begin{array}{c}
		1 \\ 0 \\ \vdots \\ 0
	\end{array} \right],\ 
x_2 = \left[ \begin{array}{c}
		0 \\ 1 \\ \vdots \\ 0
	\end{array} \right],\ \ldots\ ,\ 
x_{d_2} = \left[ \begin{array}{c}
		0 \\ 0 \\ \vdots \\ 1
	\end{array} \right] 
\end{equation}
It represents a set of fiducial effects for $\mathcal{S}_2$ since all
effects can be represented as unit vectors on the sphere. 
For any state $\psi\in \mathcal{S}_2$ we have that the probability
for the $i$-th fiducial effect is $p(x_i) = E_{i} (\phi)=
(1+ \phi^i)/2$, hence $\phi^i = 2p(x_i) - 1$. This proves the thesis.
$\blacksquare$

\subsection{Quantum theory}\label{qt}

At this point of our derivation we have not
yet considered composite systems. In this section we are going to show
that the only non classical theory satisfying axioms 1-3 and axiom
composition is quantum theory. 

First we will use axiom composition to prove what in \cite{axioh,
  axiom, axiob} is called ``axiom subspace'' and what in \cite{axioq}
is derived from ``axiom compression'' (a slighlty different
restatement of axiom subspace). In section \ref{sa} it will be discussed
the significance of that axiom and will be pointed out that it may be regarded as a mathematical
requirement on the state space of a physical system rather than a natural
informational or operational constraint.

The strategy to prove ``axiom subspace'' is to show that the field of
numbers $\mathbb{K}$ on which is defined a projective space of a
physical system satisfying axioms 1-4 does not depend on the system
considered but is a property of the theory. This will imply that the
object describing a system with information capacity $m+1$, a
$\mathbb{K}P^m$, is the same object describing a subspace with
information capacity $m+1$ of a larger system described by a
$\mathbb{K}P^n$ with $n>m$. From this fact, any representation of the
state space of the system with information capacity $m+1$ can be
equivalently considered a representation of a subspace with
information capacity $m+1$ of a larger system and this will prove
axiom subspace in our derivation. 

We will now consider two
projective spaces representing two physical systems of a theory
satisfying axioms 1-4 and the composite system obtained composing
them. Considering any of the subsystems of the composite system, a subspace of the
composite system, we will show that all projective spaces describing
systems in a theory satisfying axioms 1-4 must be defined over the
same field of numbers.       

\begin{definition}
Two projective spaces $P_1$ and $P_2$ are \emph{isomorphic} iff there exists
a bijective map between $P_1$ and $P_2$ such that the points in a
subspace of dimension $n$ of $P_1$ are mapped into points of a
subspace of dimension $n$ of $P_2$ and conversely.
\end{definition}

In the following lemma we will consider two different projective spaces
$\mathbb{K}P^{n}$ and $\mathbb{L}P^m$ describing two systems of
information capacity $n+1$ and $m+1$ respectively and the projective
space describing the system obtained composing the two systems
$\mathbb{J}P^{l}$ that must have information capacity
$l+1=(m+1)(n+1)$. We will then construct an isomorphism between
a subspace of dimension $n$ ($m$) of $\mathbb{J}P^{l}$ and
$\mathbb{K}P^{n}$ ($\mathbb{L}P^m$).

\begin{lemma}\label{field}
For any probabilistic theory satisfying axioms 1-4, given two systems
of information capacity $m+1$ and $n+1$, and the composite system
obtained composing them, there exists an isomorphism between any of
the subsystems and a subspace of the composite system with the same
information capacity.
\end{lemma}
\emph{Proof:} Suppose to have two systems $K$ and $L$ satisfying axioms 1-4 of information capacity
$n+1$ and $m+1$ respectively. According to theorem \ref{veblen} their
state spaces constitute two projective spaces of dimension $n$ and $m$ respectively defined over a skew
field. Let $K$ be represented by $\mathbb{K}P^{n}$ and $L$ by $\mathbb{L}P^m$. Composing the two systems we obtain,
from axioms 1-4, another projective
space that can be represented in $(d_n+1)(d_m+1)-1$ (excluding the
normalization degree of freedom) real
euclidean space and that we denote as $J = \mathbb{J}P^{l}$ with $l+1 =
(m+1)(n+1)$. Consider any state $\rho \in K$ and a fixed pure state $\phi
\in  L$. The map obtained as:
\begin{equation}
\phi: \rho \rightarrow \rho \otimes \phi
\end{equation} 
Is a map between states in $K$ and states in $J$. We now show that
$\phi$ is an isomorphism between $K$ and a subspace of dimension
$n$ of $J$ showing that $\mathbb{K}P^{n}$ is isomorphic to a subspace
of dimension $n$ of $\mathbb{J}P^{l}$. To see this let $\rho'$ be a
completely mixed state in $K$.
$F_{\rho'\otimes\phi}$ is, by definition, a subspace of dimension $n$
of $J$. The map $\phi$ is injective from $K$ to $F_{\rho\otimes\phi}$ and, since $\phi$ is a
fixed pure state, also surijective. This means that we have a
bijective map between pure states in $K$ and pure states in
$F_{\rho'\otimes\phi}$. Moreover, mapping any state in a
subspace of dimension $h<n$ of $K$ results a state in a subspace of
dimension $h$ of $F_{\rho'\otimes\phi}$. Thus we have an isomorphism
between a subspace of $J$ with dimension $n$, $\mathbb{J}P^{n}$, and
the $\mathbb{K}P^{n}$ describing $K$. Reversing the roles of $L$ and
$K$ in the above argument we have the thesis.   
$\blacksquare$    

\begin{theorem}\label{field1}
The field of numbers on which are defined projective spaces
representing pure states of systems in a theory satisfying axioms 1-4
is a property of the theory and does not depend on the system considered. 
\end{theorem}
\emph{Proof:} Composing a system $K$
described by $\mathbb{K}P^n$ with a system $L$ described by $\mathbb{L}P^m$, 
it results a composite system $J$ described by $\mathbb{J}P^l$ with
$l+1=(m+1)(n+1)$. From lemma \ref{field}, it exists an isomorphism
between $\mathbb{L}P^m$ and a subspace of dimension $m$ of $J$,
$\mathbb{J}P^m$ and also an isomorphism
between $\mathbb{K}P^n$ and a subspace of dimension $n$ of $J$,
$\mathbb{J}P^n$. From standard projective geometry, \cite{casse}, this
is possible only if $\mathbb{J}=\mathbb{K}=\mathbb{L}$. Since this
holds for systems of arbitrary finite information capacity we have the
thesis.
$\blacksquare$

We now turn to the last part of our derivation where we will show that
in a theory satisfying axioms 1-4, the field of numbers $\mathbb{K} = \mathbb{C}$.

From axiom composition we know that the system composed
of two elementary systems has information capacity four and we denote
the set of states of the composed system $\mathcal{S}_4$. We also know that if $d_{2} +
1$ is the number of parameters required to specify a state of an
elementary system, $(d_2+1)^2$ parameters will
suffice to specify a state of the system composed of two elementary
systems. 

In \cite{axioh} it is shown that if a theory satisfies axiom
composition then the dimension of an elementary system must be odd. Since the group of
transformations of an elementary system must be transitive on a sphere
in $\mathbb{R}^{d_2}$, it must coincide with one of the Lie
groups whose action is transitive on a sphere in $\mathbb{R}^{d_2}$
with odd $d_2$. The possibilities are: $SO(d_2)$; the smallest
exceptional Lie group usually denoted as $\mathcal{G}_2$, whose fundamental
representation is a subgroup of $SO(7)$. This observation can be found
in \cite{axiom}. We now rule out the last
possibility in that list with the following:

\begin{lemma}\label{octo}
$\mathcal{G}_2$ cannot be the group of transformations of an elementary system
in a theory satisfying axioms 1-4.
\end{lemma}
\emph{Proof:} $\mathcal{G}_2$ is the group of automorphisms of the algebra of
octonions. If this were the group of transformations of an
elementary system, then the state space of such a system would be a
projective space of dimension 1 over octonions. Consider a system
composed of two elementary systems of this kind. The state space of
the composite system must form a projective space of dimension 3 from
axiom composition and axioms 1,2. Considering one of the two subsystems as a subspace
of the composite system, the 3-dimensional
projective space should contain a one dimensional subspace which is an
octonion one dimensional projective space. But this would mean that
the 3-dimensional projective space we are dealing with is an
octonionic projective space and this contradicts theorem
\ref{veblen} since octonions are not a skew filed (they lack associativity of
multiplication). Hence we reach a contradiction and we
prove the thesis.
$\blacksquare$

The following proposition will be used in the proof of the subsequent
lemma.

\begin{proposition}\label{purprod}
A separable state $\psi$ is pure iff the marginal states on both
subsystems are pure.
\end{proposition}
\emph{Proof:} Let $\phi_1$ and $\phi_2$ be the marginal states of
$\psi$ on system A and B respectively. Suppose $\phi_1,\phi_2$ pure and $\psi$ mixed. Then
$\psi = p_a \phi'_1 \otimes \phi'_2 + p_b \phi''_1\otimes \phi''_2$ and
the marginal state on any of the subsystems would be mixed. This proves
one implication. Suppose by converse that $\psi$ is pure
and that $\phi_1$ is mixed. Then we would have $\psi = (p_A \phi'_1 +
p_B \phi''_1) \otimes \phi_2 = p_A \phi'_1 \otimes \phi_2 + p_B
\phi''_1 \otimes \phi_2$ and $\psi$ would be mixed. This proves the
thesis.
$\blacksquare$

From axiom reversibility we know that the set of transformations of
$\mathcal{S}_4$ forms a compact group and any such group can be represented by means of
orthogonal matrices. In the following lemma we will find such
representation. The argument used in the proof of the following lemma
is invented in \cite{axiom}. 

\begin{lemma}\label{rep2bits}
The representation of pure states of $\mathcal{S}_4$ in which
the group of transformations is represented by means of orthogonal matrices
is such that a pure normalized state $\psi$ is represented as follows:
\begin{equation}\label{rep2bit}
\psi= \begin{pmatrix}
\alpha_i = 2p(x_i) - 1 \\     
\vdots  \\ 
\beta_j =  2p(y_j) - 1\\   
\vdots  \\ 
\gamma_{ij} = 4p(x_i,y_j) - 2p(x_i) - 2p(y_j) +1 \\ 
\vdots  \\
1 \\
\end{pmatrix} 
\end{equation}
where $\{x_i\}_{i=1}^{d_2}$, $\{y_j\}_{j=1}^{d_2}$ are sets of fiducial effects for
$\mathcal{S}_2$, $\{\alpha_i\},\{\beta_j\}$ represent the marginal
states of $\psi$, $\{\gamma_{ij}\}$ represents the correlation matrix and the last entry
represents normalization.
\end{lemma}
\emph{Proof:} 
A normalized state $\psi \in \mathcal{S}_4$ by axiom Local Tomography
can be represented as in (\ref{rep2bit}).
By axiom Reversibilty we know that every pure state in
$\mathcal{S}_4$ can be expressed as $A \phi_1 \otimes \phi_2$ with $A$
matrix (not necessarily orthogonal) representing an element in the group of reversible transformations of
$\mathcal{S}_4$ and $\phi_1 \otimes \phi_2$ pure state represented as
in (\ref{rep2bit}). 
In the case $A$ is product we already know that $A$ is an orthogonal
matrix by lemma \ref{repbit}. Since the group of reversible transformations
in $\mathcal{S}_4$ is compact we know that it exists a
matrix $S$ such that
$S^{-1} A S = O$ with $O$ orthogonal and $S \in \mathbb{R}^{(d_2^2+
  2d_2)} \times \mathbb{R}^{(d_2^2+
  2d_2)} $ for every reversible transformation in $\mathcal{S}_4$ (not
only the product ones).
The matrix $S$ is non singular since it must perform a change of
basis from the representation (\ref{rep2bit}) to the representation in
which $A$ is an orthogonal matrix.
We will show that $S$ is
proportional to the identity, namely, those two representations are the
same. If $A$ is product then it is
orthogonal; from this we have that if $A$ is product then $A\equiv O'_1
\otimes O'_2$ and $SA =
AS$. Product transformations form a subgroup of the group of
transformations of $\mathcal{S}_4$. The following three subspaces:
i) real span of the vectors $(\alpha_1,...,\alpha_{d_2})^T$; ii) real
span of vectors $(\beta_1,...\beta_{d_2})^T$; iii) real span of matrices
$\{\gamma_{ij}\}_{i,j=1}^{d_2}$ are three invariant subspaces for the
subgroup of product transformations. By Shur's lemma we thus have:
\begin{equation}
	S=\left(\begin{array}{ccc}
		aI_{d_2} & 0 & 0  \\
		0 & bI_{d_2} &0 \\	
		0 & 0 & sI_{d_2^2}
	\end{array}\right)
\end{equation}
for some $a,b,s > 0$ where $I_{d}$ is the identity $d \times d$
matrix. Now define $\phi^0 = \theta$ and $\phi^{1} = -\theta$ with
$\theta \in \mathbb{R}^{d_2}$ and $\phi^{0,1}$ pure perfectly distinguishable states in
$\mathcal{S}_2$.
Since product of two pure states is pure, from proposition \ref{purprod} we have
$\phi^a \otimes \phi^b$ is pure  $\forall a,b \in \{0,1\}$. By axiom
Reversibility there exist transformations $G_{sw}$ and $G_{cnot}$ such
that $G_{sw} \phi^a \otimes \phi^b = \phi^b \otimes \phi^a$ and
$G_{cnot} \phi^a \otimes \phi^b = \phi^a \otimes \phi^{a\oplus
  b}$ where $\oplus$ denotes sum modulo 2. This implies that $G_{sw}
(\theta, 0, 0) = (0, \theta, 0)$ while $G_{cnot} (0,0,\theta\theta^T)
= (0,\theta,0)$ where the first and second entries in these vectors
represent vectors $\alpha$ and $\beta$, i.e. the marginal
states of the two component systems, while the third entry
represents matrix $\gamma$ containing the information regarding
correlations of the component systems (see (\ref{rep2bit})). Rewriting these two expressions in the representation
where transformations are orthogonal matrices we have: 
\[
S G_{sw} S^{-1} (a\theta, 0,0) = (0,b\theta,0)
\] 
  and
\[
S G_{cnot} S^{-1} (0,0,s\theta\theta^T) = (0,b\theta,0)
\]
From lemma \ref{repbit} we have $||\theta||^2 = 1$ and
$|| \gamma_{\phi_0\otimes \phi_0}||^2 =
\text{Tr}[\gamma_{\phi_0\otimes \phi_0}^T \gamma_{\phi_0\otimes \phi_0} ] = 1$ where
$\gamma_{\phi_0\otimes \phi_0} = \theta \theta^T$. Since $S G S^{-1} $
is orthogonal for all $G$ we have that $(a\theta, 0,0)$,
$(0,0,s\theta\theta^T)$ and $(0,b\theta,0)$ have the same modulus hence
$a=s=b$. This implies that $S=aI$ and proves the thesis.
$\blacksquare$

\begin{lemma}\label{steff}
For every pure state $\psi \in \mathcal{S}_4$ there exists an effect that gives probability
one on it.
\end{lemma}
\emph{Proof:} The thesis holds for pure product states of
$\mathcal{S}_4$ from the results obtained in subsection \ref{elsys} and the fact that probabilities for product states
and product effects factorize.
Consider performing a transformation $O$
on the composed system in a product state $\psi_{\text{prod}}$, then
performing its inverse $O^{-1}$ and, after that, making a measurement
containing the effects giving probability 1 on $\psi_{\text{prod}}$. $O$ acts
after state $\psi_{\text{prod}}$
while $O^{-1}$ acts before the effect giving probability 1 on $\psi_{\text{prod}}$ thus
transforming this effect. Expliciting this using the notation of
(\ref{probabilities0}), (\ref{probabilities}), we have: 
\begin{equation}\label{nota}
(\psi_{\text{prod}} | O^{-1}O \psi_{\text{prod}})
= (\psi_{\text{prod}} O^{-1} |O \psi_{\text{prod}}) = 1 
\end{equation}
Where $(\psi_{\text{prod}} | $ represents the effect giving
probability 1 on state $\psi_{\text{prod}}$.
By axiom reversibility, every state in $\mathcal{S}_4$ can be
written as $O\psi_{\text{prod}}$ for some reversible transformation
matrix $O$. This implies the thesis.
$\blacksquare$

\begin{definition}\label{correspo}
Given a state $\psi \in \mathcal{S}_4$, the effect giving probability one
on $\psi$ involved in lemma \ref{steff} will be called
\emph{the effect corresponding to $\psi$}. 
\end{definition}

\begin{lemma}\label{normal}
Any state vector $\psi \in \mathcal{S}_4$ in the representation of
lemma \ref{rep2bits} is such that $||\psi||^2 = 4$.
\end{lemma}
\emph{Proof:} The thesis holds by lemma \ref{repbit} for pure product
states. By axiom Reversibility and the fact that in the
representation of lemma \ref{rep2bits} transformations are represented
as orthogonal matrices the thesis follows also for every $\psi \in \mathcal{S}_4$.
$\blacksquare$

\begin{lemma}
In the representation of lemma \ref{rep2bits}, given any state $\psi$, the effect
corresponding to $\psi$ is represented by a vector proportional to
that representing $\psi$.  
\end{lemma}
\emph{Proof:} For product states the thesis holds with the
corresponding effect being a product effect. This comes from lemma
\ref{rep2bits} and the fact that probabilities for product states and
product effects factorize. In the representation of lemma
\ref{rep2bits}, since transformations are represented as orthogonal matrices, we
have: $(\psi_{\text{prod}} O^{-1} |= 1/4 \psi_{\text{prod}}^TO^T$ (the
factor 1/4 comes from normalization) while
$|O \psi_{\text{prod}})= O \psi_{\text{prod}}$ and the probability is
simply the scalar product of these two vectors.
$\blacksquare$

We already know that a two dimensional subspace of
the state space of $\mathcal{S}_4$ is a
representation of the state space of an elementary system. In the
following lemma we will show that the
set of effects corresponding to the pure states in a two dimensional
subspace of $\mathcal{S}_4$ are the set
of effects of an elementary system. We will thus prove the following:

\begin{lemma}\label{eqsubs}
The set of effects corresponding (in the sense of definition \ref{correspo}) to the pure states in a two dimensional
subspace of $\mathcal{S}_4$ forms the same manifold as the effects
associated to an elementary system. 
\end{lemma}
\emph{Proof:} From lemma \ref{ball} any two dimensional subspace
$F_{\rho}^2$ of $\mathcal{S}_4$ constitutes a sphere. From
lemma \ref{field} this
is the same manifold formed by the state space of an elementary
system $\mathcal{S}_2$, namely, a $d_{2}$-dimensional sphere with transitivity group
SO$(d_2)$ with $d_2$ odd. This is the case since both $F_{\rho}$ and
$\mathcal{S}_2$ are a projective space of dimension one over a given
field of numbers $\mathbb{K}$. Let $\mathcal{T}$ be the subset of the set of transformations
of $\mathcal{S}_4$ that transforms pure states in $F^2_{\rho}$ into
pure states in $F^2_{\rho}$. Since the manifold
representing states in $F^2_{\rho}$ is the same representing states in
$\mathcal{S}_2$, the action of $\mathcal{T}$
on these state vectors represents the group of transformations of an
elementary system. Since $\mathcal{T}$ represents such a group, we have
that also the effects
corresponding to the pure states in $F_{\rho}^2$, by definition \ref{correspo}, are represented by a
$d_2$-dimensional sphere. This implies that states in $F_{\rho}^2$ and
the effects corresponding to them are an equivalent representation of the states
and effects of an elementary system.
$\blacksquare$

In the following lemmas we will use the following notation.
\begin{itemize}
\item $(e|$ represents the deterministic effect
\item ${}_{\mathcal{S}_2}(e|\psi)_{\mathcal{S}_4}$ indicates the
marginal state of $\psi$, namely $\alpha_{\psi}$ or $\beta_{\psi}$. 
\item ${}_{\mathcal{S}_{2}}(a|\psi)_{\mathcal{S}_4}$ indicates the not
normalized  state
of $\mathcal{S}_2$ obtained measuring an effect $(a|$ on one
component system of the two bits system in state $\psi \in \mathcal{S}_4$. 
\item ${}_{\mathcal{S}_{2}}(b| \otimes
{}_{\mathcal{S}_{2}}(a| |\psi)_{\mathcal{S}_4} = (ab|\psi) $ where
$(a|$, $(b|$ are two effects in $\mathcal{S}_2$.
\end{itemize}

\begin{lemma}\label{2bitdetermin}
Given $(0|,(1|$ two perfectly discriminating effects in $\mathcal{S}_2$, the deterministic effect for the system composed of two bits is:
\begin{equation}\label{det2bit}
{}_{\mathcal{S}_4}(e| = (00| + (11| + (10| + (01|
\end{equation}
where $(ij|= (i|\otimes (j|$, $i,j= \{0,1\}$
\end{lemma}
\emph{Proof:} The effect in (\ref{det2bit}) represents the coarse graining of the outcomes
obtained in a measurement for the composite system.
$\blacksquare$

\begin{lemma}\label{psient}
Given $(0|,(1|$ two perfectly discriminating effects in $\mathcal{S}_2$, there exists a state
$\psi_{\text{ent}}\in \mathcal{S}_4$ such that:
\begin{equation}
(00|\psi_{\emph{ent}}) = (11|\psi_{\emph{ent}}) = 1/2
\end{equation}  
where $(ii|= (i|\otimes (i|$, $i= 0,1$
\end{lemma}
\emph{Proof:} If $(0|,(1|$ are two
perfectly discriminating effects there will be two pure perfectly
distinguishable states $|0),|1) \in \mathcal{S}_2$ that are perfectly
discriminated by them. Let $\rho = 1/2(|00) + |11))$ be a state of $\mathcal{S}_4$. $F_{\rho}$ is a two
dimensional subspace of $\mathcal{S}_4$. From lemma \ref{eqsubs}
the set of states in $F_{\rho}$ and the set of effects corresponding
to them are represeted by the same manifolds of the corresponding sets
of an elementary system. If we represent these two sets in real
euclidean space, then they represent one the dual of the other as in
the case of an elementary system. It then must exist a state
$\psi_{\text{ent}} \in F_{\rho}$ with the claimed property. This is
represented by a state in the equator of the sphere representing pure
states in $F_{\rho}$.
$\blacksquare$

\begin{lemma}\label{reppsient}
Using the representation of lemma \ref{rep2bits} for $\psi_{\text{ent}}$ we have
$\alpha_{\psi_{\emph{ent}}} = \beta_{\psi_{\emph{ent}}} = 0$ and
\begin{equation}
\gamma_{\psi_{\emph{ent}}} = \begin{pmatrix}
 1 &  0   \\
  0 & C  \\       
\end{pmatrix} 
\end{equation}
where $C$ is a $d_{2}-1$ dimensional real matrix.
\end{lemma}
\emph{Proof:} $\alpha_{\psi_{\emph{ent}}}$ and $\beta_{\psi_{\emph{ent}}}$ are
representations of the two marginal states
${}_{\mathcal{S}_2}(e|\psi_{\text{ent}})_{\mathcal{S}_4}$ on the two
component systems of $\mathcal{S}_4$. $|0),|1)$ are the two perfectly
distinguishable states of lemma \ref{psient}. In the representation of
lemma \ref{rep2bits}, these
are represented by two
antipodal vectors $\phi^0$,$\phi^1$ on the unit sphere in
$\mathbb{R}^{d_2}$. $(0|,(1|$ are the two corresponding
discriminating effects. We know that:
\begin{equation}
{}_{\mathcal{S}_2}(e|\psi_{\text{ent}})_{\mathcal{S}_4} =
{}_{\mathcal{S}_{2}}(0|\psi_{\text{ent}})_{\mathcal{S}_4} +
{}_{\mathcal{S}_{2}}(1|\psi_{\text{ent}})_{\mathcal{S}_4} = p|x) +
(1-p) |x')
\end{equation}
where $p|x), (1-p)|x')$ are two not normalized states in
$\mathcal{S}_2$. The state $\rho = 1/2 (|0)|0) +
|1)|1))$ is perfectly distinguishable from $|1)|0)$ with the
measurement $\{(0|(0| + (1|(1|, (1|(0| + (0|(1|\}$ where $(0|(0| +
(1|(1|$ represents the coarse graining of the corresponding outcomes. The state
$\psi_{\text{ent}} \in F_{\rho}$ is perfectly distinguishable from
$|1)|0)$ with the same measurement. This implies that 
$
 (1|(0| \psi_{\text{ent}}) = 0
$. This in turns means that $p(1|x) = 0$ that is true iff $(1|x) = 0$. $|x)$ is
a pure normalized state in $\mathcal{S}_2$ that gives probability 0 on the
effect $(1|$ and the unique normalized state in
$\mathcal{S}_2$ with this
property is $|0)$. This implies $|x) = |0)$. With
the same argument we can conclude that $(0|x')=0$ and thus $|x')
= |1)$. Since by hypothesis $(0|(0|\psi_{\text{ent}}) =
(1|(1|\psi_{\text{ent}}) = 1/2$ we must have $p=1/2$. Hence we have
${}_{\mathcal{S}_2}(e|\psi_{\text{ent}})_{\mathcal{S}_4} = 1/2(|0) +
|1))$. This implies that $\alpha_{\psi_{\emph{ent}}} = \beta_{\psi_{\emph{ent}}} =
0$ since $|0)$ and $|1)$ are antipodal vectors in the representation
we are working with. Without loss of generality we can choose the
vector representing $|0)$ equal to $x_1$ where
$x_1$ is defined in (\ref{orthon}) and $|1)$ is the antipode of
$x_1$. This is
due to the arbitrarity of the choice of $|0),|1)$ in lemma
\ref{psient}. Since from this lemma $\psi_{\text{ent}}$ is such
that:
\begin{equation}
(00|\psi_{\text{ent}}) + (11|\psi_{\text{ent}}) = 1
\end{equation} 
we conclude that $\gamma_{\psi_{\text{ent}}}^{11} = 1$ where
$\gamma_{\psi_{\text{ent}}}^{ij}$ is the $ij$-th entry of the matrix $\gamma_{\psi_{\text{ent}}}$. From the fact
that $\alpha_{\psi_{\emph{ent}}} = \beta_{\psi_{\emph{ent}}} =
0$ the action of an element of the group of product transformations on
$\psi_{\text{ent}}$ is:
\begin{equation}
O_1 \otimes O_2 \psi_{\text{ent}} = O_1\gamma_{\psi_{\text{ent}}} O_2^T
\end{equation}
with $O_i$ orthogonal matrix in SO($d_2$)
and $\gamma_{\psi_{\text{ent}}}$ $d_2$ dimensional real
matrix. Suppose now that the vector
$\{\gamma_{\psi_{\text{ent}}}^{i1}\}_{i=1}^{d_2}$ is such that it
exists $\gamma_{\psi_{\text{ent}}}^{i1} \neq 0$ for some $i\neq
1$. Then, since the group of transformations of $\mathcal{S}_2$ is
transitive on the sphere, it would exist an $O_1$ such that the vector
$O_1 (\gamma_{\psi_{\text{ent}}}^{i1} \cdots
\gamma_{\psi_{\text{ent}}}^{i{d_2}})^T$ would have an entry with modulus
greater than 1. Such entry would pertain to a state vector
representing a pure state in $\mathcal{S}_4$. The representation we
are working with, is by hypothesis the one
found in lemma \ref{rep2bits}; it is clear from (\ref{rep2bit}) that there
cannot exist entries of state vectors with modulus greater than one in
this representation. This implies that
$\gamma_{\psi_{\text{ent}}}^{i1} = 0$ for all $i\neq
1$. Repeating the argument for the vector
$\{\gamma_{\psi_{\text{ent}}}^{1i}\}_{i=1}^{d_2}$ and a transformation
$O_2^T$ we also find $\gamma_{\psi_{\text{ent}}}^{1i} = 0$ for all $i\neq
1$.
$\blacksquare$

We now have to show that the dimensionality of the state space of an
elementary system in a theory satisfying axioms 1-4 is
three. In order to do so we will use a strategy inspired to that
invented in \cite{axiob}.

We introduce the following state:
\begin{equation}
\psi_{\text{ent}}^i := R^i \otimes I \psi_{\text{ent}}
\end{equation}
$I$ is the identity in $\mathcal{S}_2$ while $R^i$ is the matrix having entries:
$R^i_{kl} = 0$ if $k \neq l$, $R^i_{kk} = 1$ if $k \neq 1, i$,
$R^i_{kk} = -1$ if $k = 1,i$. We know that $\psi_{\text{ent}}^i$ is
a state in $\mathcal{S}_4$ for all $i$
since, for all $i$, $R^i$ is in $SO(d_2)$ that coincides with the group of
reversible transformations in $\mathcal{S}_2$.

\begin{lemma}\label{ipsient}
$\psi_{\emph{ent}}^i$ and $\psi_{\emph{ent}}$ are such that
$(\psi_{\emph{ent}}^i|\psi_{\emph{ent}}) = 0$
\end{lemma}
\emph{Proof:} From lemma \ref{2bitdetermin} we know that:
\begin{equation}
(00| + (11| + (01| + (10| = (e|
\end{equation}
where $(0| = x_1$, $x_1$ is defined in (\ref{orthon}) and $(1|$ is
represented by the vector antipodal to $(0|$ in the $d_2$-dimensional
sphere representing effetcs of $\mathcal{S}_2$. This choice is the same done in
lemma \ref{reppsient}.
From lemma \ref{psient} we also have that:
\begin{equation*}
(e|\psi_{\text{ent}}) = (00 + 11|\psi_{\text{ent}}) = 1
\end{equation*}
that implies $(10|\psi_{\text{ent}}) = (01|\psi_{\text{ent}}) =
0$. Now also note that:
\begin{equation*}
(00 + 11|({R^{i}}^{-1} \otimes I) (R^i\otimes I)\psi_{\text{ent}}) = (00 + 11|\psi_{\text{ent}}) = 1
\end{equation*}
and since:
\begin{equation*}
(00 + 11|({R^{i}}^{-1} \otimes I) (R^i\otimes I)\psi_{\text{ent}}) = (10 +
01|\psi_{\text{ent}}^i) = 1  
\end{equation*} 
it follows that $(00|\psi_{\text{ent}}^i) = (11|\psi_{\text{ent}}^i) =
0$. This implies that the test $\{(00| + (11|, (01| + (10|\}$ is
perfectly discriminating for
$\psi_{\text{ent}}^i,\psi_{\text{ent}}$ hence the two states are
perfectly distinguishable. From lemma \ref{eqsubs} the set of states
and the set of effects of $F_{\rho}$ have the same representation as
geometrical object in real euclidean space as the corresponding sets
of an elementary system. This means that
$\psi_{\text{ent}}^i,\psi_{\text{ent}}$ are two antipodal points of
the sphere thus $(\psi_{\emph{ent}}^i|\psi_{\emph{ent}}) = 0$.
$\blacksquare$

\begin{theorem}\label{tre}
The dimension of the state space of an elementary system is $d_2 = 3$.
\end{theorem}
\emph{Proof:} Let $\rho = 1/2 (\psi_{\text{ent}}^i +
\psi_{\text{ent}})$. $F_{\rho}$ has information
capacity two. From lemma \ref{ipsient} and lemma \ref{eqsubs} we know that
\begin{equation}\label{eq}
(\psi_{\text{ent}}|\psi_{\text{ent}}^i) = 0
\end{equation} 
We will now explicit (\ref{eq}) using the representation of lemma
\ref{rep2bits}.
First we change notation for the matrix $\gamma_{\psi_{\text{ent}}}$:
\begin{equation}
\gamma_{\psi_{\text{ent}}} =
(\gamma_{\psi_{\text{ent}}1}^{x},\gamma_{\psi_{\text{ent}}2}^{x},
\cdots, \gamma_{\psi_{\text{ent}}{d_2}}^{x}) 
\end{equation} 
where $\gamma_{\psi_{\text{ent}}i}^{x}$ is the $i$-th column vector of
the matrix $\gamma_{\psi_{\text{ent}}} $
Note that by lemmas \ref{rep2bits} and \ref{normal} we must have
\begin{equation}\label{eq1}
||\gamma_{\psi_{\text{ent}}1}^{x}||^2 = 1
\end{equation} and
\begin{equation} \label{eq2}
\sum_{\theta=1}^{d_2} ||\gamma_{\psi_{\text{ent}}{\theta}}^{x}||^2 = 3.
\end{equation}
(\ref{eq}) can be written as
\begin{equation}
\begin{aligned}
0 = 1/4 (1 + \psi^T\psi^i) =  1/4 (1 + \text{Tr}
[\gamma_{\psi_{\text{ent}}}^T \gamma_{\psi_{\text{ent}}^i}] ) =  \;\;\;\;\;\;\;\\
= 1/4 (1 + \sum_{\theta=1}^{d_2}
||\gamma_{\psi_{\text{ent}}{\theta}}^{x}||^2 - 2
||\gamma_{\psi_{\text{ent}}i}^{x}||^2 - 2 ||\gamma_{\psi_{\text{ent}}1}^{x}||^2)
\end{aligned}
\end{equation}
This, (\ref{eq1}), (\ref{eq2}) imply that:
\begin{equation}
||\gamma_{\psi_{\text{ent}}i}^{x}||^2 =
||\gamma_{\psi_{\text{ent}}1}^{x}||^2 = 1
\end{equation}
Since by definition $i = 2,..d_2$ we must have that 
$||\gamma_{\psi_{\text{ent}}i}^{x}||^2 = 1$ for all $i$. From
(\ref{eq2}) the thesis follows.
$\blacksquare$

\begin{corollary}\label{qubit}
An elementary system of a theory satisfying axioms 1-4 is a projective space of dimension 1 over complex numbers,
i.e. a $\mathbb{C}P^1$.
\end{corollary}
\emph{Proof:} From lemma \ref{repbit} an elementary system must be a
unit $d_2$ dimensional sphere in $\mathbb{R}^{d_2}$. From theorem \ref{tre}
$d_2 = 3$ and $SO(3)$ is the group of reversible transformations
transitive on a
sphere in $\mathbb{R}^3$. It is a mathematical fact \cite{zyck} that a $\mathbb{C}P^1$ is a
manifold isomorphic to the ordinary sphere in three dimensional
euclidean space and with transitivity group $SO(3)$. 
$\blacksquare$

\begin{corollary}\label{theofin}
Pure states of a system with information capacity $n+1$ described by a probabilistic theory satisfying axioms 1-4 are
points of a $\mathbb{C}P^n$. 
\end{corollary}
\emph{Proof:} From theorem \ref{theoproj}, the state space of a system
with information capacity $n+1$ must be a projective space of
dimension $n$ over some field $\mathbb{K}$. From lemma
\ref{field} and corollary \ref{qubit}, every one dimensional subspace of this
projective space must be a $\mathbb{C}P^1$. This implies the thesis.
$\blacksquare$

\section{discussion}

\subsection{The subspace axiom}\label{sa}

In \cite{axioh, axiom, axiob}, it was assumed what we here
generically call ``subspace axiom''.  In \cite{axioq} it was assumed
axiom compression stating more or less the same thing as axiom
subspace in a way consistent with the graphical formalism invented by the
same authors.
The requirement in these axioms is that it must
exist a linear mapping between a subspace of a system and the state
space of a different system
with information capacity equal to that subspace such that the two
spaces (i.e. the state space of one system and the subspace of the
other) can be regarded as different representations of the same
mathematical object. The intuitive justification
for this axiom is that the mathematical object used to
represent a set of
states depends only on the information capacity of that set. This is clearly a
mathematical requirement on the state space of a physical system and
does not deal with information theory. We see from theorem
\ref{field1} that this property is
derived by the projective space structure of quantum theory. This
structure, in turn, is derived from axiom distinguishability and axiom
conservation that deal with storage and retrieving of
information into a physical system (see section \ref{aqt}) and do not refer to any
mathematical property that the state space of a physical system must have.   
In \cite{axioh1} it has been formulated a new
axiom \emph{sturdiness} to prove axiom subspace in the duotensor
framework. Despite its "operational" formulation, \emph{sturdiness}
strongly relies on the notion of filter as used in quantum
mechanics. This axiom can thus be hardly regarded as an information
theoretic constraint on a probabilistic theory and in the context of
the present reconstruction it would sound strange. 

Requiring axioms like the subspace one in a derivation of quantum
theory from informational/operational requirements means to require a
priori much of the mathematical structure of the theory wihtout giving
a justification in terms of more basic principles. In the present
reconstruction we are able to get rid of axiom subspace in favor of
two simpler axioms related to storing and retrieving of information
into physical systems. It is remarkable that deriving an important
piece of mathematical structure of quantum theory from more basic
principles permits us to classify probabilistic models different from
quantum theory that share with quantum theory only some of the natural
requirements we imposed in section \ref{aqt} but not others.

\subsection{Complex amplitudes and composite Systems}

It is shown in section \ref{mqt} that if a probabilistic
theory satisfies axioms 1-5 then either is classical or it constitutes a
generalization of quantum theory in which the superposition
principle holds with amplitudes not necessarily complex but belonging to
a generic field of numbers. Among these theories we find
quantum theory but also quantum theory over reals and quantum
theory over quaternions corresponding respectively to amplitudes in superpositions
being reals and quaternions. The result in corollary \ref{theofin} implies
that among the theories in this class,
using complex amplitudes in superpositions is equivalent to require
axiom composition
for the description of composite systems.

It is shown in \cite{Wootershardy} that quantum
theory over reals satisfies 2-Local Tomography for the composite system.
\begin{definition}
A theory satisfies 2-Local Tomography if the state of a composite
system (ABC...) can be
determined from the statistics of measurements on the single
components (A, B, C,..) and the statistics of bipartite measurements on two
components at one time (AB, AC, BC, ...). 
\end{definition}   
2-local tomography differs from local tomography because in the latter
the statistics of measurements on the single components suffice to
determine the state. The result of \cite{Wootershardy} combined with
the present derivation of quantum theory is very interesting. We have
infact proved that quantum theory over reals is, with quantum
theory, in the
class of probabilistic theories satisfying axioms 1-5. On the other
hand quantum theory is the only theory in this class satisfying axiom
composition with local
tomography. In \cite{Wootershardy} is shown that
quantum theory over reals satisfies 2-local tomography in place of
local tomography. This implies that quantum theory over reals satisfies the same informational axioms of quantum theory
except from the one concerning composite systems in which local tomography must be
substituted with 2-local tomography. We are now in a
position to state a conjecture similar to the one stated in
\cite{Wootershardy}: quantum theory over reals is the only
probabilistic theory satisfying the following list of axioms:
system, distinguishability, capacity, conservation, reversibility and
composition with 2-local tomography.

It is very natural that our description of the physical world involves local tomography in
spite of 2-local tomography for describing state of a composite
systems. Indeed if the latter were used then the mere fact that
two different physical systems are described at once as a single
system would imply for the
composite system to have more degrees of freedom than the cartesian
product of the degrees of freedom of the components.

\section{conclusion}

The mathematical rules governing quantum theory can be understood in
terms of information theoretic principles. A physical system is
thought as something in which can be stored information. The maximal
amount of information that can be stored in a given system has a
definite value. Probabilistic theories are a very general framework in
which physical systems can be described in terms of operational
primitives such as preparations, transformations and measurement outcomes. 
Five natural axioms for physical systems in the
probabilistic theories framework are
Distinguishability, Conservation, Reversibility, Composition.
Theories satisfying the first two axioms are either classical or such that pure
states of a physical system are points of a projective space defined
over a generic field of numbers. This latter class of theories constitutes a
generalization of quantum theory in which superposition principle
holds with amplitudes not necessarily complex but belonging to a
generic field of numbers. Examples of theories in this class are
quantum theory, real quantum theory, quaternionic quantum theory. This
result establishes a connection between two different approaches to
quantum foundations, quantum logic and the one based on information
theoretic primitives. Such connection is provided by theorem
\ref{theoproj}. More importantly, this result permits to get rid of
the subspace axiom in the reconstruction. Axiom subspace was always
more or less explicitly used in previous reconstructions and contains
a mathematical constraint on the state space of a physical system. 
They are also classified all the possible probabilistic theories that are
consistent with three subsets of the set of axioms given and this
could be useful in experimental tests of quantum mechanics to check
whether the results of these tests could be consistent with other
mathematical models.

{\bf Acknowledgement} I thank P. Perinotti and A. Tosini for having
taught me much of the background needed to write this paper. I thank L. Masanes for
very useful correspondence. I thank S. Al Safi for very helpful
comments and suggestions. I also thank P. Goyal and H. Imai for very useful
critics and comments. Many thanks to the referee of Physical Review
A for a surprisingly detailed review of the paper and for unvaluable
cirtics and suggestions. I am in debt with M. Mueller that
gave me a precious insight for the proof of theorem \ref{field1} and also
with L. Hardy for very useful discussions.

\end{document}